\documentclass[%
reprint,
unsortedaddress,
showpacs,preprintnumbers,
footinbib,
amsmath,amssymb,
aps,
prl,
]{revtex4-1}

\usepackage{graphicx}
\usepackage{dcolumn}
\usepackage{bm}
\usepackage{braket}
\usepackage{gensymb}
\usepackage[usenames]{color}
\usepackage{hyperref}

\begin{document}

\preprint{UdeA-Uniandes--QOptics}

\title{Experimental verification of a quantumness criterion for single systems based on a Bell-like inequality}

\author{J. Castrill\'on,}
\email{jhonny.castrillon@udea.edu.co}
\affiliation{Instituto de F\'{i}sica, Universidad de Antioquia UdeA, Calle 70 No. 52-21, Medell\'{i}n, Colombia}

\author{O. Calder\'on-Losada, D.A. Guzm\'an, Alejandra Valencia,}
\affiliation{Quantum Optics Laboratory, Universidad de los Andes, A.A. 4976, Bogot\'a D.C., Colombia}

\author{B. A. Rodr\'iguez}
\affiliation{Instituto de F\'{i}sica, Universidad de Antioquia UdeA, Calle 70 No. 52-21, Medell\'{i}n, Colombia}

\begin{abstract}
In this letter, we propose and experimentally test a quantumness criterion for single systems. The criterion is based on the violation of an already reported classical inequality. This inequality is expressed in terms of joint probabilities that we identify as realizations  of preparation-and-measurement schemes. Our analysis points out that superposition is the key notion to grasp the quantum nature of physical systems. We verify our criterion in an all-optical experimental setup with heralded single photons generated via spontaneous parametric down conversion and use exclusively projective preparations and projective measurements.

\end{abstract}

\maketitle

\textit{Introduction}.-- Since the early days of quantum mechanics, it has been clear the necessity of \emph{quantum criteria} as a way to differentiate whether a physical system can be considered classical or not \cite{epr,anti-epr,gato}. This search has  not been free of controversy \cite{FreireBook}. Some of the proposed quantum criteria are based on the violation of inequalities as, Bell theorem that applies for spatially separated entities \citep{Bell64} or Legget-Garg inequality (LGI) that applies for successive measurements on single systems \cite{PhysRevLett.54.857}. Another criterion, the Alicki-Van Ryn, is deduced for expectation values of pairs of non-commuting observables of single systems \cite{alicki08}. In all these criteria, the authors used assumptions such as realism, locality, non-invasive measurability, macrorealism and other ideas motivated by the intuition taken from the behavior of the macroscopic world. In the literature, there are several experimental realizations with different physical systems that have demonstrated the validity of each of those criteria \cite{PhysRevLett.49.1804,0034-4885-77-1-016001,PhysRevA.79.044102}.

Any general statement about the classical/quantum frontier has to deal with the issue of whether a single system reveals quantum features or not. In this paper, we precisely addressed this issue by proposing and testing a quantum criterion for two-level single systems that, differently from the previously reported criteria, reinforces the idea that the key concept when dealing with quantumness is superposition. Our criterion is based on an inequality, that we will refer to as CISS, standing for Classical Inequality for Single Systems. This inequality is expressed in terms of joint probabilities and constitutes a simpler and more general result than Bell and Legget-Garg inequalities since it was previously found as an intermediate result by Wigner~\cite{Wigner70}, when deducing Bell inequality, and by Lapiedra~\cite{Lapiedra}, when deriving a variation of the LGI. In the present work, the joint probabilities that appear in the CISS will be interpreted in terms of the basic operations of \emph{preparation} and \emph{measurement}. This approach allows to get rid of possible ambiguities since these operations are well defined from a theoretical \cite{ballentine} and  an instrumental point of view \cite{peres} as can be recognized by the use of a generic ``prepare-and-measure'' (P\&M) scheme at different research agendas related with quantum mechanics\cite{peres,Pegg,Park1992,arXiv1509.03641,Naruse2015,PhysRevLett.112.140407,Oreshkov2012, NewJPhys17_111002}. We test our criterion in an all-optical experimental setup with heralded single photons, generated via spontaneous parametric down-conversion (SPDC) and use exclusively projective preparations and projective measurements. Additionally, in the supplementary material we present a derivation of the CISS, that instead of using the assumptions of joint reality and perfect correlations, as Lapiedra does, considers the existence of a \emph{generalized classical state}, $R$, that joints all the properties of a system and avoids the use of other ambiguous classical assumptions.

\begin{figure}[tb]
\includegraphics[width=0.5\textwidth]{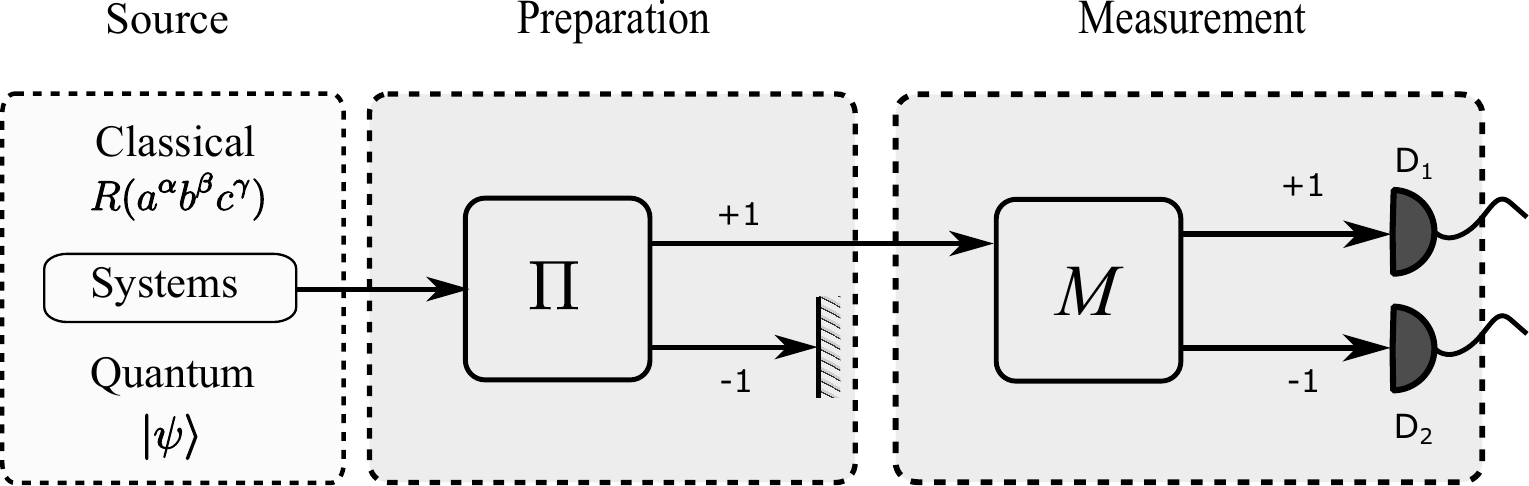}
\caption{Cartoon representing the preparation and measurement stages in a P\&M scheme for either classical or quantum domains. Systems are prepared in the $+1$ value of the property $\Pi$ and then measured in the property $M$. Detectors D$_1$ and D$_2$ collect the data for measurement of values $+1$ and $-1$ of property $M$.}
\label{cartoon}
\end{figure}

\textit{CISS and the P\&M Scheme.} -- Let us assume the existence of an ensemble of individual systems, each of them characterized by a generalized physical state $R=(a^{\alpha}b^{\beta}c^{\gamma})$. This state is a representation of the physical situation in which three dichotomous properties, $a$, $b$ and $c$, have one of its two possible outcomes $\alpha=\pm1, \beta=\pm1$, and $\gamma=\pm1$, respectively. The properties of $R$ can be related using a function $P(\pi^i \mu^j)$ that represents the joint probability that a single system has the outcome $\pi^i$ for the $\Pi$ property and the outcome $\mu^j$ for the $M$ property, regardless the value of the third one. Different joint probabilities can  be related to write a classical inequality for single systems. In particular, for the state $R=(a^+b^{-}c^{-})$, it can be demonstrated that when property  $b$ is relaxed, $P(a^+ c^-)$ must satisfy
\begin{equation}
\label{ciss}
  P(a^+ c^-)\leq P(a^+ b^-)+P(b^+ c^-),
\end{equation}
which is precisely the inequality that we call CISS. This equation sets a bound to the distribution of properties in an arbitrary ensemble of single systems and we use it to define our quantum criterion. To see this in a clearer way, let us introduce the parameter $S$ defined by
\begin{equation}
\label{S}
 S\equiv P(a^+ b^-)+P(b^+ c^-)-P(a^+ c^-).
\end{equation}
In the classical scenario in which the CISS is valid, $S\geq0$; while in the case in which the physical system is described by quantum states, it can be found that  $S<0$ as we will see in what follows.  As a consequence of the fact that the behavior of $S$ depends on considering classical or quantum systems, we associate $S$ to our quantumness criterion.

Joint probabilities of the form $P(\pi^{i} \mu^{j})$, that appear on Eq.~(\ref{S}), can be seen from the perspective of a P\&M scheme in which a system is prepared in $\pi^{i}$, a possible outcome for a property $\Pi$, and then measured in $\mu^{j}$, a possible outcome of another property $M$. Typically, in a P\&M scheme, the ``preparation'' stage is a set of temporal ordered interactions aimed to bring the system of interest to a specific state and the ``measurement'' stage is a set of operations performed on the state in order to deliver an outcome. Fig.~\ref{cartoon} represents a situation in which a system, not yet specified if it is quantum or classical, passes through a P\&M scheme in which  $\Pi$ and $M$ are dichotomic properties with possible outcomes $+1$ and $-1$. In particular, for the depicted case, the state is prepared in $\pi^+$ and measured either in $\mu^+$ by detector D$_1$ or in $\mu^-$ by detector D$_2$.

For the generalized classical state $R$, $P(\pi^i \mu^j)=P(\mu^j \pi^i)$. This is not surprising given that classically it is irrelevant the order of the preparation and measurement stages of a P\&M scheme. This means that preparing $\pi^i$ and measuring $\mu^j$ ($\pi^i\rightarrow \mu^j$) or conversely, preparing $\mu^j$ and measuring $\pi^i$ ($\mu^j\rightarrow \pi^i$) leads to the same value of their corresponding joint probabilities. This independence of the order for P\&M leads to the conclusion that the bound given by Eq.~\eqref{ciss} has to be a consequence of choosing a generalized classical state as $R$ that then defines, not just what a single system is, but how the populations, or property densities, are distributed in a classical ensemble.

\textit{Quantum superposition and behavior of the $S$ parameter.} --   In a quantum context, a dichotomic property is an observable represented by an operator, $\hat q$, with eigenvalues $q={\pm1}$ and corresponding eigenstates $\ket{q \pm}$. These properties can be incompatible. When this is the case, the outcomes of the measurement stage depend on the previously prepared state. Therefore, differently from the classical scenario, in a P\&M scheme the order of the preparation and measurement operations is relevant. For instance, in the situation shown in the Fig.~\ref{cartoon}, $\pi^+\rightarrow \mu^-\nLeftrightarrow \mu^-\rightarrow \pi^+$. This fact implies that $P(\pi^+ \mu^-)\neq P( \mu^- \pi^+)$ evoking a quantum discord, in the terms used by Zurek in the context of mutual information: ``two classically identical expressions (...) generally differ when the systems involved are quantum''~\cite{PhysRevLett.88.017901}. This fact can be clearly seen considering the case in which the physical system is an ensemble of single photons, each of them characterized by the quantum state, $\ket{\psi}$. The quantum properties, in this scenario, can be different linear polarizations. To say that a photon is in state $\ket{q+}$ means that it has a polarization oriented at $\theta_q$ with respect to the horizontal (w.r.t.h), and to say that a photon is in $\ket{q-}$ means that its polarization is oriented at $\theta_{q}^{\perp}=\theta_q + 90\degree$. In the canonical $\{\ket{H},\ket{V}\}$ basis, in which $\ket{H}$ denotes horizontal and $\ket{V}$ vertical polarization,  $\ket{q\pm}$ can be written as \cite{Beck2012},
\begin{subequations}
\label{eq:statesq}
\begin{equation}
 \ket{q+}=\cos\theta_q\ket{H}+\sin\theta_q\ket{V}\label{q+}
\end{equation}
\begin{equation}
 \ket{q-}=\sin\theta_q\ket{H}-\cos\theta_q\ket{V}.\label{q-}
\end{equation}
\end{subequations}

With these definitions, it is possible to calculate conditional probabilities as projective operations $P(\pi^i|\psi)=|\braket{\pi^i|\psi}|^2$ and $P(\mu^j|\pi^i)=|\braket{\mu^j|\pi^i}|^2$. Considering the situation of Fig.~\ref{cartoon}, $P(\mu^-|\pi^+)=|\braket{\mu-|\pi+}|^2=\sin^2(\theta_{\mu}-\theta_\pi)$ and without loosing generality, one can choose $\ket{\psi}=\ket{H}$, in such a way that the marginal probability, $P(\pi^+)$, follows $P(\pi^+)\equiv P(\pi^+|H)=|\braket{\pi+|H}|^2=\cos^2\theta_\pi$.

In general joint probabilities satisfy Bayes' rule, $P(\pi^i \mu^j)=P(\mu^j|\pi^i) P(\pi^i)$, in which joint probabilities are related with conditional  and marginal probabilities.  In light of the P\&M scheme, Bayes' rule can be seen as the product of the probability of preparing a system in $\pi^i$, $P(\pi^i)$, times the probability of measuring the state $\mu^j$ given that the system was prepared in $\pi^i$, $P(\mu^j|\pi^i)$. Invoking Bayes' rule,
\begin{eqnarray}
 P(\pi^+ \mu^-)&=&P(\mu^-|\pi^+) P(\pi^+)  \nonumber \\
 &=&\sin^2(\theta_{\mu}-\theta_{\pi})\cos^2\theta_{\pi}
\label{p&m}
\end{eqnarray}
and,
\begin{eqnarray}
  P(\mu^{-} \pi^{+})&=&P(\pi^+|\mu^-) P(\mu^-)  \nonumber\\
  &=&\sin^2(\theta_{\mu}-\theta_\pi)\sin^2\theta_\mu    \label{p&mmupi}
\end{eqnarray}
since $P(\mu^-)\equiv P(\mu^-|H)=|\braket{\mu^-|H}|^2=\sin^2\theta_\mu$. Equations (\ref{p&m}) and (\ref{p&mmupi}) yield $P(\pi^+ \mu^-)\neq P(\mu^{-} \pi^{+})$. Interestingly, the origin of this ``discord'' must be a consequence of the superposition that exist in the definitions of the quantum states in Eq.~\eqref{q+} and Eq.~\eqref{q-}, given that in our discussion we are considering single systems and there are no entangled or product states.

To see that for a quantum system, $S<0$, we choose as properties three different orientations of polarization for the three dichotomic properties, $\hat a$, $\hat b$ and $\hat c$. In this situation, $S$ becomes a function of the orientation angles of the polarizations, $S=S(\theta_a,\theta_b,\theta_c)$. Using Eq.~\eqref{p&m}, the joint probabilities in Eq.~\eqref{S} become
\begin{subequations}
\begin{equation}
  P(a^+ c^-)=\sin^2(\theta_c-\theta_a)\cos^2\theta_a, \label{ac}
\end{equation}
\begin{equation}
 P(a^+ b^-)=\sin^2(\theta_b-\theta_a)\cos^2\theta_a,\label{ab}
\end{equation}
\begin{equation}
 P(b^+ c^-)=\sin^2(\theta_c-\theta_b)\cos^2\theta_b\label{bc}.
\end{equation}
\end{subequations}

By combining these expressions in Eq.~\eqref{S}, it is found that $S_{\text{min}}=-0.403$, for $\theta_a\simeq 157\degree,\theta_b\simeq 123.5\degree$ and $\theta_c\simeq 77.5\degree$.

\begin{figure}[b]
\includegraphics[width=0.5\textwidth]{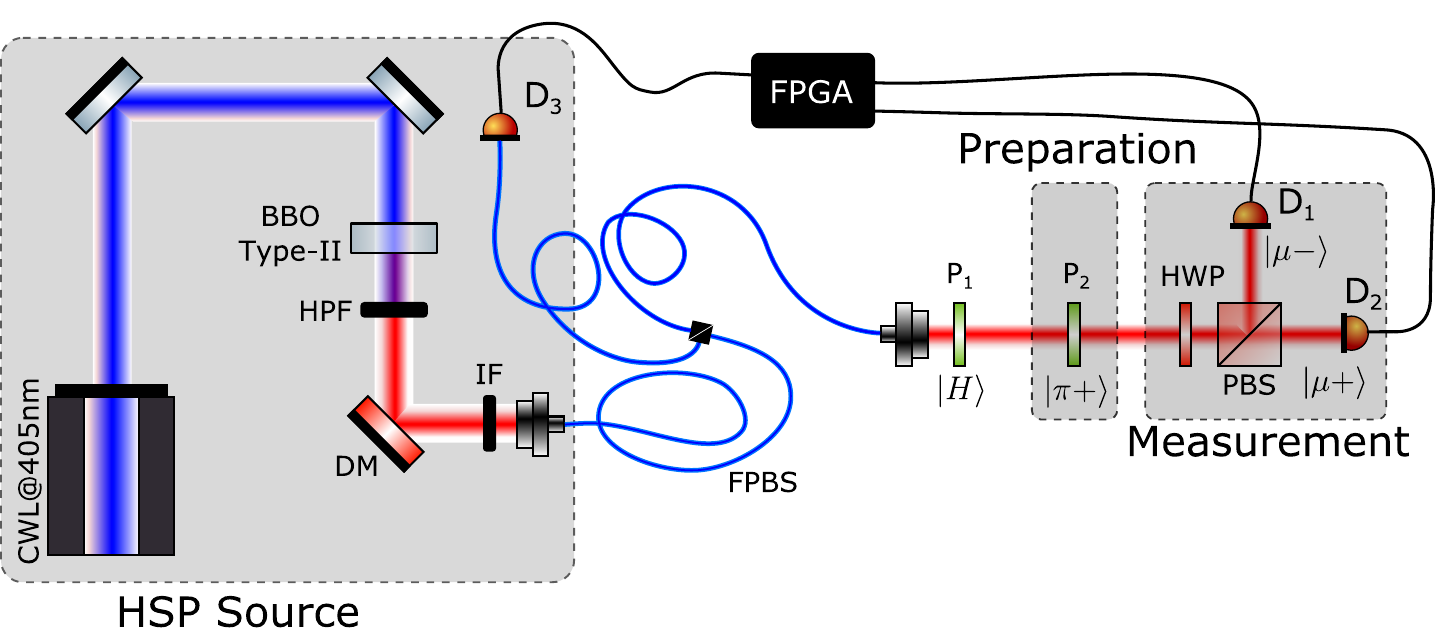}
\caption{Experimental setup. A diode laser centered at 405nm was used to pump a Type-II BBO crystal cut to produce collinear down-converted pairs of photons. A high pass filter (HPF) and a dichroic mirror (DM) remove the residual pump beam. The down-converted photons are filtered by an 810nm centered, 10nm bandwidth bandpass interference filter (IF). To use the orthogonally-polarized pairs of photons as a heralded single photon source, photons are coupled into a fiber optic polarizing beam splitter (FPBS) with polarization-maintaining fiber pigtails, obtaining one photon per output fiber. The vertically polarized photon goes directly to a single photon detector (D$_3$), while the horizontally polarized photon passes through the preparation and measurement stages, composed by a polarizer (P$_2$), a half waveplate (HWP), a polarizing beam splitter cube (PBS) and single photon detectors (D$_1$, D$_2$).}
\label{setup}
\end{figure}

\textit{Experimental setup.--} In order to test that the parameter $S$ can indeed take negative values when considering a quantum scenario, it is necessary to implement the proper P\&M scheme for each joint probability in Eq.~\eqref{S}. These three joint probabilities are equivalent in the sense that they consist on the preparation of the eigenstate with the positive eigenvalue, for the first property, and the measurement of the state with the negative eigenvalue for the second one, which is precisely the case depicted for detector D$_2$ in Fig.~\ref{cartoon}. We take advantage of this equivalence and use the experimental setup depicted in Fig.~\ref{setup}. The whole experiment can be analyzed along three main stages: (i) \textit{source:} heralded single photons (HSP) with horizontal polarization $\ket{H}$ are generated via type II SPDC \cite{Shih2003}. The HSP source is based on the fact that in the SPDC process two photons, known as signal and idler, are generated at the same time and therefore one of the photons can be used to announce the presence of its twin, that will play the role of the single photon during the experiment. For this setup, we used a 0.5 mm BBO crystal in a collinear configuration, pumped by a CW laser at 405 nm. The output from the crystal was coupled to a fiber polarizing beam splitter (OZ Optics) whose fibers were single mode and polarization maintaining. The vertical polarized photon was used as trigger when detected by an avalanche photo-diode (SPCM-AQRH-13-FC), labeled as D$_3$ in Fig.~\ref{setup}. To further ensure the correct polarization of the HSP, polarizer P$_1$ was used in front of the fiber output; (ii) \textit{preparation:} using polarizer P$_2$, the different properties $\hat{\pi}$ are prepared in its eigenstate $\ket{\pi+}$ defined by the angle $\theta_{\pi}$  w.r.t.h.; and (iii) \textit{measurement:} a half-wave plate (HWP) followed by a polarizing beam splitter (PBS) works like a linear polarization analyzer which outcomes define $\ket{\mu+}$ and $\ket{\mu-}$, the eigenstates of the property $\hat{\mu}$. Detector D$_1$ counts how many of the photons, prepared in the state $\ket{\pi+}$, are measured in state $\ket{\mu-}$, and detector D$_2$ counts how many of the photons prepared in state $\ket{\pi+}$ are measured in the state  $\ket{\mu+}$. Therefore, by counting coincidences D$_1$\&D$_3$ and D$_2$\&D$_3$, one can obtain a measurement of $P(\pi^+\mu^-)$ and $P(\pi^+\mu^+)$, respectively. The photodetectors outputs are analyzed by means of a Field Programable Gate Array (FPGA) set to count singles and coincidences in a 9 ns window.

The angles associated to $\hat{\pi}$ and $\hat{\mu}$ properties are set by orientation of P$_2$ ($\theta_{\pi}$) and the HWP ($\theta_{\mu}$), respectively. In particular, $\theta_{\pi}$ and $\theta_{\mu}$ were varied in the range $[0\degree,180\degree]$ in steps of $6\degree$. In the actual implementation, the HWP was scanned between $[0\degree,90\degree]$ in steps of $3\degree$ due to its working principle. For all the combinations of these angles, the probabilities $P(a^+c^-)$, $P(a^+b^-)$ and $P(b^+c^-)$ from equations \eqref{ac}--\eqref{bc} are obtained in terms of $\theta_a$, $\theta_b$ and $\theta_c$, and the parameter $S(\theta_a,\theta_b,\theta_c)$ is reconstructed.

\begin{figure}[t]
\includegraphics[width=0.45\textwidth]{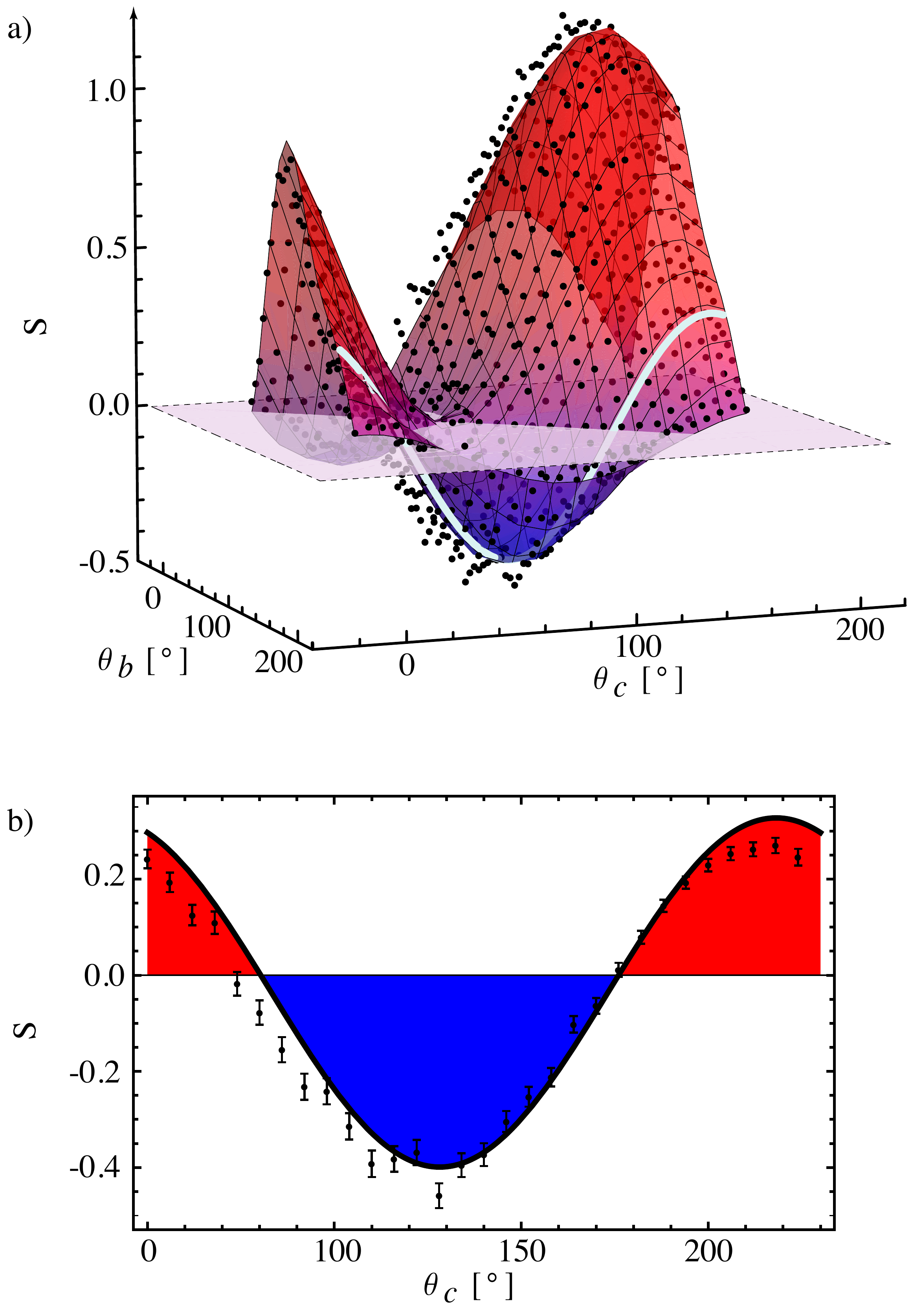}
\caption{Theoretical prediction and experimental data for a) $S(\theta_a=156\degree,\theta_b,\theta_c)$ and b) $S(\theta_a=156\degree,\theta_b=126\degree,\theta_c)$. The fact that $S$ can reach negative values is clearly observed.}
\label{Srecos}
\end{figure}

Figure~\ref{Srecos}a  depicts the theoretical surface, $S(\theta_a=156\degree,\theta_b,\theta_c)$, and  the experimental dots for $S$ when $\theta_a^{\text{exp}} = 156\degree$. This value of $\theta_a$ is chosen because in our experiment, it is the closer that we can get to the theoretical prediction  $\theta_a\simeq 157\degree$ that minimizes $S$.
The plane $S=0$ serves to underline the negative non-classical region of the $S$ parameter. In Fig.~\ref{Srecos} b, we depict a profile of $S$, by choosing $\theta_b$ as the nearest experimental angle to the theoretical predicted value for maximum violation that can be reach with our experimental apparatus, $S(\theta_a=156\degree,\theta_b=126\degree,\theta_c)$. The agreement between theory and experiment is evident. The minimum experimental value of $S$ is $S_{\text{min}}^{\text{exp}}=-0.39\pm 0.03$ that violates CISS by $17$ standard deviations.

\textit{Discussion and Conclusion.--} We have reported and experimentally verified a quantum criterion for single systems. In this work, we interpreted joint probabilities on the light of a P\&M scheme. The validity of this interpretation can be seen by the agreement between our experimental data and the theoretical prediction in Fig.~\ref{Srecos}. The use of a P\&M scheme removes possible controversial issues and allow us to conclude that the key notion when dealing with quantumness is superposition.

The criterion we proposed is based on a previously derived inequality, that we called CISS, and the definition of a parameter $S$. We proved the validity of our criterion by finding $S<0$ in an all-optical experiment that uses heralded single photons and projective preparation and projective measurement. Traditionally, this type of violation is seen as a no-go theorem implying, in our case, that the generalized classical state $R$ used in the derivation of the CISS, is \emph{no} longer useful in the quantum realm. By assuming $R$ for quantum systems it is possible to obtain $S< 0$ which is meaningless from the classical point of view.

To find that $S< 0$ together with the fact that the CISS is based on the existence of the generalized classical state, and therefore no reference needs to be made to the concept of nonlocality, reinforces the idea that the key notion when dealing with quantumness is superposition. We do not discuss whether nonlocality exists or not. We only claim that there is no need to think in nonlocality to grasp quantumness: On a fundamental level, all is about superpositions as written in Eq.~\eqref{q+} and Eq.~\eqref{q-}.  With this in mind, we can say that what experiments testing violations of Bell inequality do for entanglement, the experiment here reported does for superposition: It indicates that \textit{superposition is quantum}.


\begin{acknowledgments}
\textit{Acknowledgments.} The authors acknowledge partial financial support from Facultad de Ciencias, Universidad de los Andes, Bogot\'a, Colombia, from Departamento Administrativo de Ciencia, Tecnolog\'\i a e Innovaci\'on (COLCIENCIAS), and from Universidad de Antioquia, under Projects No. 2014-989 (CODI-UdeA) and Estrategia de Sostenibilidad del Grupo de F\'\i sica At\'omica y Molecular. They also thank Juan Pablo Restrepo Cuartas for helping with the design of the graphs.
\end{acknowledgments}

\bibliography{evQcbli}

\end{document}